\documentclass[aps,prd, twocolumn,groupedaddress,amsmath,amssymb,preprintnumbers,nofootinbib]{revtex4-2}

\usepackage[dvipdfmx]{graphicx}
\usepackage{dcolumn}
\usepackage{bm}
\usepackage{braket}
\usepackage{breqn}
\usepackage{xcolor}
\usepackage{hyperref}

\definecolor{myblue}{RGB}{0, 100, 200}

\begin{document}


\title{Decoherence of the quantum superradiant cloud from graviton emission}

\author{Lingyun Fu}
\email{l46fu@uwaterloo.ca}
\affiliation{Department of Applied Mathematics and Theoretical Physics$,$ University of Cambridge$,$\\Wilberforce Road$,$Cambridge$,$ CB3 0WA$,$ UK}

\author{Hidetoshi Omiya}
\email{omiya@tap.scphys.kyoto-u.ac.jp}
\affiliation{Department of Physics$,$ Kyoto University$,$ Kyoto 606-8502$,$ Japan} 

\author{Takahiro Tanaka}
\email{t.tanaka@tap.scphys.kyoto-u.ac.jp}
\affiliation{Department of Physics$,$ Kyoto University$,$ Kyoto 606-8502$,$ Japan}
\affiliation{Center for Gravitational Physics and Quantum Information$,$ Yukawa Institute for Theoretical Physics$,$ Kyoto University$,$ Kyoto 606-8502$,$ Japan}

\author{Hui-Yu Zhu}
\email{hzhuav@ibs.re.kr}
\affiliation{Cosmology$,$ Gravity and Astroparticle Physics Group$,$ Center for Theoretical Physics of the Universe$,$ Institute for Basic Science (IBS)$,$ Daejeon$,$ 34126$,$ Korea}

\date{\today}

\begin{abstract}
Black hole superradiance amplifies vacuum fluctuations into a macroscopic cloud of a massive bosonic field around a rotating black hole. 
Under free evolution, the quantum cloud is a two-mode squeezed state that remains entangled even at large occupation numbers, while its growing component admits a classical statistical description. 
In this work, we study how graviton emission changes the quantum state of the cloud. We find that graviton emission suppresses interference between cloud configurations with different amplitudes, but leaves interference between configurations that differ only by an overall sign unchanged. For typical configurations realized under vacuum fluctuations, the suppression becomes significant when the expected number of emitted gravitons is of order unity.We also show that, at larger occupation numbers, all first and second moments can be reproduced by a statistical mixture of coherent states.
For the astrophysical black holes, both changes occur much before the spin-down of the black hole becomes significant.
\end{abstract}


\maketitle

\section{Introduction}

Black hole superradiance, a process in which bosonic waves extract energy from a rotating black hole~\cite{Press:1972zz}, provides a promising way to search for ultralight bosonic fields~\cite{Arvanitaki:2009fg,Arvanitaki:2010sy,Brito:2015oca}.
For a massive bosonic field, 
the amplified modes can remain gravitationally bound to the black hole, repeatedly undergoing superradiant amplification, leading to an instability~\cite{Damour:1976kh,Zouros:1979iw,Detweiler:1980uk,Dolan:2007mj}. 
The instability results in the spontaneous formation of a macroscopic cloud that can potentially produce various observational signatures.
Such observable signatures are the characteristic black hole spin distributions~\cite{Brito:2014wla,Stott:2018opm,Ng:2019jsx,Hoof:2024quk}, gravitational waves emitted from the cloud with frequencies can fall within the sensitivity bands of current and future gravitational wave detectors~\cite{Brito:2017zvb,Brito:2017wnc,Tsukada:2018mbp,Isi:2018pzk,Ghosh:2018gaw,Zhu:2020tht,Siemonsen:2022yyf,LISAConsortiumWaveformWorkingGroup:2023arg,Omiya:2024xlz} , modifications to binary dynamics and stellar orbits~\cite{Baumann:2018vus,Baumann:2019ztm,Ding:2020bnl,Takahashi:2021yhy,Tong:2021whq,Tong:2022bbl,Baumann:2021fkf,Cole:2022yzw,Takahashi:2023flk,Tomaselli:2023ysb,Fan:2023jjj,GRAVITY:2023cjt,Spieksma:2024voy,Zhu:2024bqs,Tomaselli:2024dbw,Ding:2025nxe,Dyson:2025dlj,Tomaselli:2025jfo,Ding:2025hqf,Dyson:2025dlj}, and imprints on photon ring observations~\cite{Chen:2022nbb,Chen:2022kzv}.

The cloud is generally assumed to grow from quantum fluctuations and then amplified by the superradiant instability. Previous studies expect that a classical field description becomes a good approximation at large occupation number~\cite{Arvanitaki:2010sy,Arvanitaki:2014wva}.
However, large occupation alone does not justify the classical description. 
Interactions with the environment have been suggested as a possible mechanism~\cite{Baryakhtar:2020gao}, but no explicit description of this process was given in that work.

In Ref.~\cite{Fu:2025ztk}, we studied the quantum production of the superradiant cloud.  Quantizing a massive scalar field on Kerr space-time, we show that the superradiant instability can be regarded as a spontaneous pair production of the positive- and negative-energy particles. This pair creation process would lead to exponential growth of the particle number. Furthermore, the state of the cloud can be described as the two-mode squeezed state, which exhibits the entanglement between the positive and negative modes. The mechanism here is similar to the case of the primordial perturbation generated during inflation~\cite{Grishchuk:1989ss, Polarski:1995jg}. Under free evolution, the two modes remain entangled even when their occupation numbers become large.

As the cloud evolves toward the macroscopic regime, the cloud emit gravitons which can carry information about the cloud and act as an environment. The emission of the graviton can lead to decoherence which may contribute to the emergence of the classical behavior of the cloud by suppressing the quantum interference between the distinct cloud configuration. The decoherence induced by the graviton emission has been well-studied for superpositions of massive particles~\cite{Oniga:2017pyq,DeLisle:2019dyw}. The problem we would like to answer is when the decoherence due to graviton emission becomes important and how the quantum correlation of the cloud is changed.

In this paper, we study the evolution of the quantum superradiant cloud  under the effect of the graviton. 
We first clarify that the growing component of the cloud admits a classical statistical description even in the absence of decoherence. However, the full state of the cloud remains coherent and entangled. 
To determine how graviton emission modifies these quantum properties, we derive the master equation of the density matrix of the cloud by tracing out the graviton environment. We find that the interference between the cloud with different amplitude would be significantly suppressed. For typical configurations of the cloud seeded by the vacuum fluctuation, the suppression becomes significant when the expected number of emitted gravitons exceeds order unity.
At a larger occupation number, we show that all first and second moments of the cloud can be reproduced by a classical mixture of coherent states. Therefore, measurements of the first and second moments alone cannot distinguish the cloud from such a classical mixture. Furthermore, in the case of astrophysical black holes, the loss of the quantum coherence occurs much before the spin-down saturation of the central black hole.

This paper is organized as follows.
In Sec.~\ref{sec:2}, we review the quantum description of the superradiant cloud and discuss its quantum properties.
In Sec.~\ref{sec:3}, we derive the interaction between the cloud and gravitons and relate the interaction strength to the gravitational wave energy flux.
In Sec.~\ref{sec:4}, we derive the master equation and estimate the scales at which decoherence becomes important.
Section~\ref{sec:5} summarizes the paper.
We use units $c=G=\hbar=1$ throughout the paper.

\section{Quantum superradiant cloud}
\label{sec:2}

\subsection{Quantization of superradiant mode}

We first summarize the quantum description of the superradiant cloud developed in Ref.~\cite{Fu:2025ztk}. 
We consider a free massive scalar field $\phi$ on Kerr spacetime. The action is given by
\begin{align}
	S &= \int d^4 x \, \sqrt{-g} \left(-\frac{1}{2}g^{\mu\nu}\partial_\mu \phi \partial_\nu\phi - \frac{1}{2}\mu^2 \phi^2\right)~,
\end{align}
where $g_{\mu\nu}$ denotes the Kerr metric.  The mode functions of the scalar field include the continuum modes and the discrete superradiant mode. The superradiant mode is labeled by the integers $(n,l,m)$. Below, we focus on one superradiant mode.

\begin{figure}[t]
\centering
\includegraphics[keepaspectratio, scale=1.0]{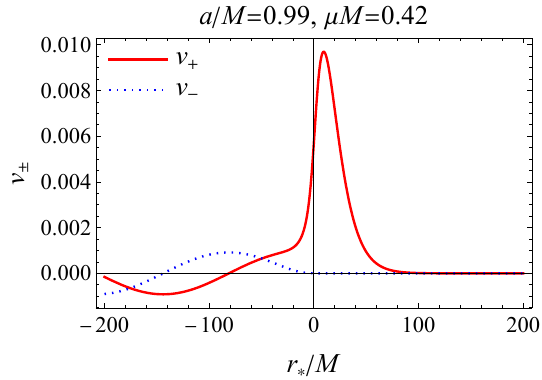}
\caption{Radial profiles of the mode functions $v_+$ (red solid curves) and $v_-$ (blue dotted curves) at $t=0$, $\varphi=0$, and $\theta=\pi/2$. We plot the mode with $n=2,l=m=1$. The real and imaginary parts are shown for $v_+$ and $v_-$, respectively. We set $a/M=0.99$ and $\mu M=0.42$.}
\label{fig:modefnc}
\end{figure}

After neglecting the modes other than the superradiant mode of interest, the field operator is represented as
\begin{align}\label{eq:modeexp0}
	\hat{\phi}(x) &=  v_{l m n, +}(x) \hat{a}_{l m n, +} + \bar{v}_{l m n, +}(x) \hat{a}_{l m n, +}^\dagger \cr
    &\qquad + v_{l m n, -}(x) \hat{a}_{l m n, -} + \bar{v}_{l m n, -}(x) \hat{a}_{l m n, -}^\dagger ~,
\end{align}
where $v_{\pm}$ corresponds to the linear combination of the mode function of the unstable superradiant mode and its decaying counterpart. The explicit expressions for the mode functions are 
\begin{align}
	v_{l m n, +}(t,\bm{x}) &= \frac{1}{\sqrt{2}}e^{-i (\omega_{l m n, R} t -m \varphi )} \cr
    &\qquad \times \biggl(e^{\omega_{l m n, I} t} R_{n l m}(r) S_{n l m}(\theta)\cr
    &\qquad \qquad + e^{-\omega_{l m n, I} t} \bar{R}_{n l m}(r) \bar{S}_{n l m}(\theta)\biggl)~,\cr
	v_{l m n, -}(t,\bm{x}) &= \frac{1}{\sqrt{2}}e^{+i (\omega_{l m n, R} t -m \varphi )} \cr
    &\qquad \times \biggl(  e^{\omega_{l m n, I} t} \bar{R}_{n l m}(r) \bar{S}_{n l m}(\theta) \cr
    &\qquad\qquad -e^{-\omega_{l m n, I} t} R_{n l m}(r) S_{n l m}(\theta)\biggl)~,\label{eq:modefuncpm}
\end{align}
where we assume both $\omega_{lmn,R}$ and $\omega_{lmn,I}$ are positive. 
The functions $R_{n l m}$ and $S_{n l m}$ are the solutions of
\begin{align}
		&\frac{d}{dr}\left(\Delta\frac{dR_{n l m}}{dr}\right) \cr
        &\qquad + \left[\frac{K^2(\omega)}{\Delta} - \mu^2 r^2 -\lambda_{n l m}(\omega_{l m n}) \right]R_{n l m} = 0~,\label{eq:EOMrad}
\end{align}
\begin{align}
	&\frac{1}{\sin\theta}\frac{d}{d\theta}\left(\sin\theta \frac{d{S}_{n l m}}{d\theta}\right) \cr
    &\qquad + \left[c^2(\omega_{l m n}) \cos^2 \theta - \frac{m^2}{\sin^2\theta}\right]S_{n l m} = - \Lambda_{n l m}S_{n l m}~,\cr\label{eq:EOMang}
\end{align}
where
\begin{align}
	\Delta &= (r- r_+)(r- r_-) \cr 
    K(\omega) &= (r^2+a^2)\omega - am~,\cr
	 c^2 (\omega) &= a^2 (\omega^2 - \mu^2)~, \cr 
     \lambda_{l m}(\omega) & = -2am \omega +a^2\omega^2 +\Lambda_{n l m}~.
\end{align}
Here, $r_+ (r_-)$ denotes the outer (inner) horizon, and $a$ is the spin of the black hole.
The eigenvalue $\Lambda_{lmn}$ and the complex frequency $\omega_{lmn}=\omega_{lmn, R}+i\omega_{lmn, I}$ are fixed by the following boundary conditions on $S_{nlm}$ and $R_{nlm}$. For the angular function $S_{nlm}$, we impose regularity at $\theta = 0,$ and $\pi$. The radial function $R_{nlm}$ must be ingoing at the future horizon and exponentially decaying at infinity, that is
\begin{align}
	R_{nlm} &\to \frac{\mathcal{N}}{\sqrt{r^2 + a^2}}\begin{cases}
	A^{\rm in}_{nlm} e^{-i k_H r_*}~, & (r_* \to -\infty)~,\\
	e^{+i k_\infty r_*}~, & (r_* \to +\infty)~,
	\end{cases}
\end{align}
with $k_H = \omega_{nlm} - m \Omega_H$, and $k_\infty = \sqrt{\omega_{nlm}^2 - \mu^2}$. The angular velocity of the black hole horizon is denoted as $\Omega_H = a/(r_+^2 + a^2)$ and the tortoise coordinate $r_*$  is given by $dr_*/dr = (r^2 + a^2)/\Delta$.
The normalization factor $\mathcal{N}$ is chosen so that the Klein-Gordon norm of $v_\pm$ is unity. Under our normalization, the energy of the scalar field $\phi = v_+ + \bar{v}_+$ on $t = {\rm const.}$ time slice becomes $\sim \omega_{nlm,R}$. The mode function $v_{+}(t,\bm{x})$ ($v_{-}(t,\bm{x})$) is localized mainly outside (inside) the potential barrier for $|\omega_I t| \ll 1$ (see Fig.~\ref{fig:modefnc}). Below, we abbreviate the labels $n,l,m$ to simplify the notation. 

Substituting the field operator into the energy-momentum tensor, we obtain the Hamiltonian of the superradiant mode as
\begin{align}\label{eq:hamiltonian}
	\hat{H}_{0} &= \hat{H}_R + \hat{H}_S~,\cr
	\hat{H}_R &= \omega_{R}(\hat{a}_{+}^\dagger \hat{a}_+ - \hat{a}_{-}^\dagger \hat{a}_-)~, \cr
	\hat{H}_S  &=  i \omega_I (\hat{a}_+^\dagger \hat{a}_-^\dagger - \hat{a}_+ \hat{a}_-)~,
\end{align}
up to an additive constant. The Hamiltonian shows that the superradiant instability can be regarded as two-mode squeezing, which has also been shown for superradiant scattering~\cite{Matacz:1992gtm}.
Since the sign in front of - mode in $\hat H_R$ is negative, we can regard the - mode as the negative energy particle. 
The time evolution under $\hat{H}_0$ is
\begin{align}\label{eq:BT}
	\begin{pmatrix}
	\hat{a}_+(t_0)\\
	\hat{a}_-^\dagger(t_0)
	\end{pmatrix}
	 &= e^{-i\omega_R t_0}
	 \begin{pmatrix}
	 \cosh\omega_I t_0 &  \sinh\omega_I t_0\\
	 \sinh\omega_I t_0 & \cosh\omega_I t_0
	 \end{pmatrix}
	\begin{pmatrix}
	\hat{a}_+\\
	\hat{a}_-^\dagger
	\end{pmatrix}~.
\end{align}

\subsection{Properties of unstable vacuum}

The unstable superradiant mode has no stationary ground state. For this reason, we consider a vacuum defined at a chosen reference time~\cite{PhysRev.57.315,PhysRevD.14.1939, Kang:1997uw}. Expanding the field in the modes $v_\pm(t-t_0,\bm{x})$ defines the annihilation operators $\hat a_\pm(t_0)$. The corresponding vacuum depending on the reference time $t_0$ satisfies
\begin{align}
	\hat{a}_+(t_0)\ket{\Omega(t_0)} &= \hat{a}_-(t_0)\ket{\Omega(t_0)} =0~.
\end{align} 
Operators at different reference times are related by the Bogoliubov transformation~\eqref{eq:BT}.

We take $\ket{\Omega(0)}$ as the initial state of the cloud. In the occupation number basis at $t=t_0$, the vacuum state can be written as
\begin{align}\label{eq:TMSS}
	\ket{\Omega(0)} &= \frac{1}{\cosh\omega_I t_0}\sum_{n=0}^{+\infty} \left(\tanh \omega_I t_0\right)^n \ket{n,n}_{t_0}~,\\
	\ket{n_+,n_-}_{t_0} &= \frac{1}{\sqrt{n_+! n_- !}}\left(\hat{a}_+^\dagger(t_0)\right)^{n_+} \left(\hat{a}_-^\dagger(t_0)\right)^{n_-}\ket{\Omega(t_0)}~.
\end{align}
The expression shows the entanglment of this state between the $+$ and $-$ modes. 
The mean occupation numbers of the instantaneous modes at $t=t_0$ are
\begin{align}\label{eq:numberpm}
	N &=\braket{\Omega(0)|\hat{a}_+^\dagger(t_0)\hat{a}_+(t_0)|\Omega(0)} \cr
    &= \braket{\Omega(0)|\hat{a}_-^\dagger(t_0)\hat{a}_-(t_0)|\Omega(0)}  = \sinh^2\omega_I t_0~.
\end{align}
The two modes also have the correlation
\begin{align}\label{eq:corrpm}
	M = \braket{\Omega(0)|\hat{a}_+(t_0)\hat{a}_-(t_0)|\Omega(0)} &= \frac{\sinh 2\omega_I t_0}{2}~.
\end{align}

Up to here, we considered in the Heisenberg picture. For the computation in Sec.~\ref{sec:4}, the Schr\"{o}dinger picture is convenient. 
The state at time $t$ is given by
\begin{align}
	\ket{\psi_S(t)} &= \exp\left[-i \hat{H}_{0} t\right]\ket{\Omega(0)}~.
\end{align}
We stress that $\ket{\psi_S(t)} $ is not same as $\ket{\Omega(t)}$. The density matrix of the cloud in the Schr\"{o}dinger picture is given by
\begin{align}\label{eq:freecloudsch}
	\rho_S(t) &= \ket{\psi_S(t)} \bra{\psi_S(t)}~.
\end{align}

As seen by the entanglement between the $+$ and $-$ modes, the quantum cloud has a distinct feature from the classical cloud. Let us quantify the entanglement in more detail. Since the first moments $\langle a_\pm\rangle$ vanish and the vacuum remain Gaussian under the evolution with the quadratic Hamiltonian, the state is fully characterized by its covariance matrix.  We introduce the Schr\"odinger operator of the quadrature vector
\begin{align}
	\hat{\bm R}
	&:=
	(\hat{x}_+,\hat{x}_-,\hat{p}_+,\hat{p}_-)^{\mathrm T}~,\\
	\hat{x}_\pm
	&:=\frac{\hat a_\pm + \hat a_\pm^\dagger}{\sqrt{2}}~,
	&
	\hat p_\pm
	&:=\frac{\hat a_\pm - \hat a_\pm^\dagger}{\sqrt{2}i}~,
\end{align}
and define the covariance matrix by
\begin{align}	\label{eq:covdef}
	V_{ij}
	&=
	\frac{1}{2}
	\braket{
	\Delta\hat R_i\Delta\hat R_j
	+
	\Delta\hat R_j\Delta\hat R_i}~,
\end{align}
where $\Delta\hat R_i=\hat R_i-\braket{\hat R_i}$. 
Here, the expectation value $\braket{\cdot}$ is taken with the operators at $t = 0$, such as $N(t) = {\rm Tr}\left[\hat{a}^\dagger_+ \hat{a}_+ \rho_S(t)\right]$. 
The covariance matrix of any physical state satisfies the uncertainty relation with the symplectic form $[R_i,R_j]=i\Omega_{ij}$, namely
\begin{align}\label{eq:uncertainty}
	V+\frac{i}{2}\Omega &\ge 0~,
&
	\Omega=
	\begin{pmatrix}
	0&0&1&0\\
	0&0&0&1\\
	-1&0&0&0\\
	0&-1&0&0
	\end{pmatrix}~.
\end{align}
Here the matrix inequality means that $V+i\Omega/2$ is positive semidefinite, that is
\begin{align}
	\bar{\bm{v}}\cdot\left( V + \frac{i}{2} \Omega \right)  \bm{v} \ge 0
\end{align}
holds for any $\bm{v}$.

The sufficient condition for the state to be entangled in terms of the second-order moments can be obtained through the partial transposition of the covariance matrix~\cite{Duan:2000awi,Simon:2000zz,Weedbrook:2011wxo}. Partial transposition is an operation on a system with two modes that transposes the density matrix with respect to only one of them. Suppose that the state is not entangled, that is, the density matrix of the system can be written in the separable form
\begin{align}\label{eq:separable}
	\rho = \sum_i p_i \rho_{+,i} \otimes \rho_{-,i}~,
\end{align}
where $0 \le p_i \le 1$ and $\sum_i p_i = 1$. Let us transpose $-$ mode.  The transposed density matrix $\rho_{-,i}^{T} = \bar{\rho}_{-,i}$ is still another physical density matrix. Therefore, the partial transposed covariance matrix must also satisfy the uncertainty relation \eqref{eq:uncertainty}. In terms of the covariance matrix, the partial transposition corresponds to flipping the sign of $\hat{p}_-$. Writing the resulting covariance matrix as $\widetilde{V}$, we obtain
\begin{align}\label{eq:PTseparable}
	\widetilde{V}+\frac{i}{2}\Omega &\geq 0~.
\end{align}
Violation of Eq.~\eqref{eq:PTseparable} certifies entanglement for an arbitrary state. For a two-mode Gaussian state, Eq.~\eqref{eq:PTseparable} is sufficient and necessary condition for separability.

Let us apply the criteria to the state of the superradiant cloud under free evolution~\eqref{eq:freecloudsch}. For the state~\eqref{eq:freecloudsch}, the first moments $\braket{\hat a_\pm}$ and $\braket{\hat a^\dag_\pm}$ vanish and the following correlations also vanish:
\begin{align}
	\braket{\hat a_\pm^2}
	&=
	\braket{\hat a_+^\dagger\hat a_-}
	=0~,
&
	{\rm Im}\braket{\hat a_+\hat a_-}&=0~.
\end{align}
Note that this is true even if we include the graviton emission.
Thus, the covariance matrix takes the form
\begin{align}\label{eq:covblock}
	V
	&=
	\begin{pmatrix}
	A&C&0&0\\
	C&B&0&0\\
	0&0&A&-C\\
	0&0&-C&B
	\end{pmatrix}~,
\end{align}
where
\begin{align}
	A&=\braket{\hat a_+^\dagger\hat a_+}+\frac{1}{2} = N + \frac{1}{2}~,\cr
	B&=\braket{\hat a_-^\dagger\hat a_-}+\frac{1}{2} = N + \frac{1}{2}~,\cr
	C&=\operatorname{Re}\braket{\hat a_+\hat a_-} = M~.
\end{align}
After partial transposition, we obtain
\begin{align}\label{eq:PTblock}
	\widetilde{V}
	&=
	\begin{pmatrix}
	K&0\\
	0&K
	\end{pmatrix}~,
&
	K&=
	\begin{pmatrix}
	A&C\\
	C&B
	\end{pmatrix}~.
\end{align}
Let $S$ diagonalize $K$ as $S^T K S = {\rm diag}(\tilde{\nu}_+, \tilde{\nu}_-)$, with the eigenvalues
\begin{align}\label{eq:PPTnuABC}
	\tilde{\nu}_\pm
	&=
	\frac{1}{2}
	\left[
	A+B\pm\sqrt{(A-B)^2+4C^2}
	\right]~.
\end{align}
After rotating the quadrature by the orthogonal matrix
\begin{align}
    T&:=
	\begin{pmatrix}
	S&0\\
	0&S
	\end{pmatrix} ~ , 
\end{align}
together with $T^T\Omega T= \Omega$
, we have
\begin{align}\label{eq:PTcovariance}
	\widetilde{V} + \frac{i}{2}\Omega  = 
	\begin{pmatrix}
	\tilde{\nu}_+&0&i/2&0\\
	0&\tilde{\nu}_-&0&i/2\\
	-i/2&0&\tilde{\nu}_+&0\\
	0&-i/2&0&\tilde{\nu}_-
	\end{pmatrix}~,
\end{align}
The eigenvalues of the matrix~\eqref{eq:PTcovariance} is $\tilde{\nu}_{+} \pm 1/2$ and $\tilde{\nu}_- \pm 1/2$. Since the smallest eigenvalue is $\tilde{\nu}_- - 1/2$, the uncertainty relation~\eqref{eq:PTseparable} reduces to the condition
\begin{align}\label{eq:PPTbound}
	\tilde{\nu}_-&\geq \frac{1}{2}~.
\end{align}
However, in the case of our present two-mode squeezed state, we have
\begin{align}\label{eq:PPTnuTMSS}
	\widetilde\nu_-
	&=
	N+\frac12-M
	=
	\frac12e^{-2\omega_I t_0}~.
\end{align}
Thus, the bound~\eqref{eq:PPTbound} is violated for all $t_0>0$. The two modes therefore remain entangled even when a large occupation number is realized.

For later convenience, we introduce the combinations
\begin{align}\label{eq:growdecay}
	\hat{c}_g
	&=\hat{a}_+ + \hat{a}_-^\dagger~, &
	\hat{c}_d
	&=\hat{a}_+ - \hat{a}_-^\dagger~.
\end{align}
From the Bogoliubov transformation~\eqref{eq:BT}, their free Heisenberg evolution is
\begin{align}
	\hat c_g(t_0)&=e^{- i \omega_R t_0 + \omega_I t_0}\hat c_g~,
&
	\hat c_d(t_0)&=e^{- i \omega_R t_0 -\omega_I t_0}\hat c_d~,
\end{align}
where the subscripts $g$ and $d$ denote growth and decay. The operators satisfy
\begin{align}\label{eq:commutationc}
	[\hat c_g,\hat c_g^\dagger]
	&=[\hat c_d,\hat c_d^\dagger]=0~,
&
	[\hat c_g,\hat c_d^\dagger]&=2~.
\end{align}

To calculate the covariance matrix during graviton emission, we use the three second moments
\begin{align}\label{eq:M2G2Qdef}
	G_2(t)&=\braket{\hat{c}_g^\dagger\hat{c}_g}~,\\
	M_2(t)&=\braket{\hat{c}_d^\dagger\hat{c}_d}~,\\
	Q(t)
	&=\braket{
	\hat{a}_+^\dagger\hat{a}_+
	-
	\hat{a}_-^\dagger\hat{a}_-}
	=\frac12\braket{
	\hat{c}_g^\dagger\hat{c}_d
	+
	\hat{c}_d^\dagger\hat{c}_g} + 1~.
\end{align}
Their relation to $A$, $B$, and $C$ in Eq.~\eqref{eq:covblock} is
\begin{align}\label{G2M2def}
	G_2&=A+B+2C~,\cr
	M_2&=A+B-2C~, \cr
	Q &= A - B~.
\end{align}
Consequently, the eigenvalue $\widetilde\nu_-$ can be written as
\begin{align}\label{eq:PPTnu}
	\widetilde\nu_-
	&=
	\frac14
	\left[
	G_2+M_2
	-
	\sqrt{(G_2-M_2)^2+4Q^2}
	\right]~.
\end{align}
For the freely evolving two-mode squeezed state,
\begin{align}
	G_2(t_0)&=e^{2\omega_I t_0}~,
&
	M_2(t_0)&=e^{-2\omega_I t_0}~,
&
	Q(t_0)&=0~.
\end{align}
Note that since the superradiant instability is pair creation process the difference of the occupation numbers between $+$ and $-$ modes, $Q$, is conserved for free evolution. 
However, graviton emission generates $Q\neq0$, so that all three moments are needed to evaluate Eq.~\eqref{eq:PPTnu}.

For the calculation of decoherence in Sec.~\ref{sec:decoherence}, the eigenstates of $\hat{c}_g$ will give convenient basis. Since $[\hat{c}_g,\hat{c}_g^\dagger]=0$, we can introduce the eigenstate which diagonalize $c_g$ and $c_g^\dagger$ simultaneously. We label the state by a complex number $c$,
\begin{align}\label{eq:cgeigenbasis}
    \hat{c}_g\ket{c}&=c\ket{c}~,
&
    \braket{c'|c}&=\delta^{(2)}(c-c')~.
\end{align}
Here, $c$ is the eigenvalue of $\hat c_g$ at $t=0$ and represents the initial amplitude of the growing mode. Note that the states $\ket{c}$ are not ordinary coherent states. Since 
\begin{align}
	\hat{c}_g &= \frac{\hat{x}_+ + \hat{x}_- + i \left(\hat{p}_+ - \hat{p}_-\right)}{\sqrt{2}}~,
\end{align}
the state $\ket{c}$ is simultaneous eigenstate of the position $\hat{x}_+ + \hat{x}_-$ and the momentum $\hat{p}_+ - \hat{p}_-$. 

Focusing on the growing component neglecting the decaying one also gives a ``classical" description of the cloud. First of all, since $\hat{c}_g$ and $\hat{c}_g^\dagger$ commute, once we focus only on the growing mode, $\hat{c}_g$ can be treated as if it were a complex random variable. 
The classical amplitude of the growing mode evolves as
\begin{align}\label{eq:classicalgrowth}
	 \dot c_{\rm cl}(t)&= (\omega_I - i \omega_R) c_{\rm cl}(t)~,
\end{align}
with its initial condition drawn from the probability distribution~\footnote{We use the fact that the vacuum is a Gaussian state.}
\begin{align}\label{eq:cgdistribution}
    p_0(c_{\rm cl}(0))&=|\braket{c_{\rm cl}(0)|\Omega(0)}|^2
    =\frac{1}{\pi}e^{-|c_{\rm cl}(0)|^2}~.
\end{align}
Averaging over this amplitude distribution reproduces all moments involving only the growing-mode components. 
In this sense, the growing component of the cloud admits a classical statistical description regardless of the decoherence.  A similar classical description is used for the growing mode of inflationary perturbations assuming that the decaying mode can be neglected~\cite{Polarski:1995jg}. 

The classical statistical description above does not imply that interference between two states with different growing-mode amplitudes is suppressed. For the pure initial vacuum, we have the amplitude of the interference between them is simply evaluated as 
\begin{align}
\left|\langle c|\hat{\rho}_S(0)|c'\rangle\right|=\left|\langle c|\Omega(0)\rangle \langle\Omega(0)|c'\rangle\right|
=\sqrt{p_0(c)p_0(c')}~.
\end{align}
Observables depending only on the growing mode probe the diagonal distribution and are insensitive to these off-diagonal elements because of the orthogonality \eqref{eq:cgeigenbasis}. However, the correlations involving $\hat{c}_d$ are sensitive to the off-diagonal coherences. For example, the full two-mode covariance matrix, which includes $\hat{c}_d$, certifies entanglement during free growth. We next examine how graviton emission affects the off-diagonal coherences in the growing mode basis and the entanglement signature encoded in the second moments.

\section{Interaction with graviton}
\label{sec:3}

We now consider the coupling of the scalar cloud to the graviton. To linear order in the metric perturbation $\hat \psi_{\mu \nu}$, the interaction Hamiltonian is
\begin{align}
	\hat H_{\rm int} &= -\frac{1}{2}\int d^3 \bm{x}' \sqrt{-g(x')}\hat{\psi}_{\mu\nu}(x')\hat T^{\mu\nu}~.
\end{align}
We focus on the emission of the graviton from the cloud. For this purpose, we only keep the oscillating component of the energy-momentum tensor. By decomposing the terms into the growing and decaying combinations~\eqref{eq:growdecay}, we have
\begin{align}\label{eq:Ttimedep}
	\hat T_{\mu\nu}^{\rm osc} &= \frac{1}{2}\left( t_{\mu\nu}^{(g)} \hat c_g^2 + 2 t^{(s)}_{\mu\nu}\hat c_g\hat c_d + t^{(d)}_{\mu\nu}\hat c_d^2 + ({\rm h.c.})\right)~.
\end{align}
Note we abbreviate the terms in $T_{\mu\nu}$ that is proportional to the background Kerr metric $g_{\mu\nu}$, since they do not contribute in the radiation gauge used below. 
Here, 
\begin{align}
	t^{(g)}_{\mu\nu} &= \frac{1}{2}\left[\partial_{\mu} (v_+ + \bar{v}_-) \right]\left[\partial_{\nu} (v_+ + \bar{v}_-)\right] ~, \cr
    t^{(s)}_{\mu\nu} &= \frac{1}{2}\left[\partial_{(\mu} (v_+ + \bar{v}_-)  \right]\left[\partial_{\nu)} (v_+ - \bar{v}_-)\right]~, \cr
	 t^{(d)}_{\mu\nu} &= \frac{1}{2}\left[\partial_{\mu} (v_+ - \bar{v}_-) \right]\left[\partial_{\nu} (v_+ - \bar{v}_-)\right] ~.  & &
\end{align}
The linear combinations of the mode functions $v_+ \pm \bar{v}_-$ can be written as
\begin{align}
	v_+ + \bar{v}_- &= \sqrt{2} e^{- i (\omega_R t -m \varphi )} e^{+ \omega_I t} R S~, \cr
	v_+ - \bar{v}_- &= \sqrt{2} e^{- i (\omega_R t -m \varphi )} e^{-\omega_I t} \bar{R} \bar{S}
\end{align}
which display the growing and decaying time dependences explicitly. 

The graviton operator $\hat{\psi}$ is expanded as~\cite{Candelas:1981zv}
\begin{align}
	\hat{\psi}_{\mu\nu} &= \sum_{l,m,P}\int d\omega \left(\overrightarrow{\Pi}_{\mu\nu}(x;l m \omega P)\hat{b}^{({\rm up})}_{lm\omega P} \right.\cr
    &\qquad \left.+ \overleftarrow{\Pi}_{\mu\nu}(x;l m \omega P)\hat{b}^{({\rm in})}_{lm\omega P} + {\rm (h.c.)}\right)~.
\end{align}
The up and in mode functions $\overrightarrow{\Pi}$ and $\overleftarrow{\Pi}$ vanish on $\mathcal{I}^-$ and $\mathcal{H}^-$, respectively~\footnote{We use $\overrightarrow{\Pi}$ and $\overleftarrow{\Pi}$ to label the mode function of up- and in-mode. These functions have different normalization from the usual one.}. They can be written in terms of Teukolsky functions using metric reconstruction~\cite{Chrzanowski:1975wv}. For example, in the ingoing radiation gauge, $\overleftarrow{\Pi}$ can be represented as
\begin{align}
	\overleftarrow{\Pi}_{\mu\nu}&(x;l m \omega \pm) \cr
    & = {}_{+2}\tau^*_{\mu\nu}\left[{}_{-2}\overleftarrow{R}_{lm\omega}(r) {}_{+2}S_{lm\omega}(x) \frac{e^{-i(\omega t - m \varphi)}}{\sqrt{2\pi}}\right]\cr
	&\qquad \pm {}_{+2}\overline{\tau^*}_{\mu\nu}\left[{}_{-2}\overleftarrow{R}_{lm\omega}(r) {}_{-2}S_{lm\omega}(x) \frac{e^{-i(\omega t - m \varphi)}}{\sqrt{2\pi}}\right]~.\cr
\end{align}
The differential operator ${}_{+2}\tau_{\mu\nu}^*$ is given, for example, in App.~A of Ref.~\cite{Sago:2005fn}. The boundary condition for ${}_{-2}\overleftarrow{R}_{lm\omega}(r)$ is given by
\begin{align}
	{}_{-2}\overleftarrow{R}_{lm\omega}(r) \to
	\begin{cases}
		{}_{-2}\overleftarrow{A} \Delta^2 e^{-i k_H r_*}~, & (r_* \to -\infty)~,\\
		\frac{1}{\sqrt{4\pi \omega} (-2 i \omega)^2}\frac{e^{-i \omega r_*}}{ r} \\ \qquad + {}_{-2}\overleftarrow{B}r^3 e^{+i \omega r_*}~, & (r_* \to +\infty)~.
	\end{cases}
\end{align}

In the following analysis, we focus on the interaction with the in-mode, in which its coupling with the cloud is related to the flux to infinity. The up-mode contribution associated with horizon flux is omitted. The size of the horizon flux is not evaluated here, however we expect it to be smaller than the flux to infinity since the dominant part of the cloud reside outside of the potential barrier (see Fig.~\ref{fig:modefnc}). The interaction Hamiltonian in the interaction picture is then approximated by
\begin{align}\label{eq:inthamilt}
	\hat H_{\rm int} &\sim - \frac{1}{4}\sum_{l,m,P}\int_{0}^{+\infty}d\omega \cr
    &\qquad \times \left\{ \hat{b}^{{\rm (in)}\dagger}_{lm\omega P} \hat{J}_{lm\omega P} e^{+i(\omega -\omega_{\rm gw})t} + {\rm h.c.}\right\}~,
\end{align}
where we introduce the emission operator 
\begin{align}\label{eq:fullsource}
	\hat{J}_{lm\omega P} &= C^{(g)}_{lm\omega P} e^{+ 2 \omega_I t} \hat{c}_g^2 + C^{(s)}_{lm\omega P}  (\hat{c}_g \hat{c}_d + \hat{c}_d \hat{c}_g)  \cr
    &\qquad  + C^{(d)}_{lm\omega P} e^{- 2 \omega_I t} \hat{c}_d^2~.
\end{align}
The frequency and azimuthal quantum number of the emitted graviton are related to those of the cloud $\omega_c$ and $m_c$ by $\omega_{\rm gw} = 2 \omega_{c,R}$ and $m_{\rm gw}=2m_c$, respectively. The factor of two arises because the energy-momentum tensor is quadratic in the scalar field $\phi$. 
The coefficient $C^{(g)}_{lm\omega P}$ is defined by
\begin{align}\label{eq:coefCg}
	\int d^3 x \sqrt{-g}\, &\bar{\overleftarrow{\Pi}}_{\mu\nu}(x;l m \omega P)  t^{(g)\mu\nu} \cr
    &=  C^{(g)}_{lm\omega P} \delta_{mm_{\rm gw}} e^{+ i(\omega - \omega_{\rm gw})t + 2\omega_I t}~.
\end{align}
The coefficients $C^{(s)}$ and $C^{(d)}$ are defined analogously. The interaction considered here corresponds to the pair annihilation of a scalar into a graviton, conventionally considered in the literature~\cite{Arvanitaki:2010sy, Yang:2023vwm}. 

We can relate the coefficients to the energy flux of the gravitational waves emitted from the cloud. Direct computation shows the relation
\begin{align}\label{eq:relationCZ}
	&C^{(g)}_{lm\omega P } = \frac{(-1)^m}{i \pi (2\omega)^{3/2}}{}_{-2}Z_{lm_{\rm gw}\omega_{\rm gw}}\left(1 + P (-1)^l\right)~,\cr
	&{}_{-2}Z_{lm_{\rm gw}\omega_{\rm gw}} = \cr 
    &\qquad 2i (4\pi \omega)^{3/2}\int dr dx (r^2 + a^2 x^2)\Delta^{-2} \cr
	&\qquad \qquad  \times {}_{-2}\overleftarrow{R}_{l m_{\rm gw} \omega_{\rm gw}}(r) {}_{-2}S_{lm_{\rm gw}\omega_{\rm gw}}(x) {}_{-2}T_{lm_{\rm gw}\omega_{\rm gw}}(r,x)~.\cr
\end{align}
where ${}_{-2}T_{lm_{\rm gw} \omega_{\rm gw}}$ is the source term of the Teukolsky equation given by
\begin{align}
	 {}_{-2}T_{lm_{\rm gw}\omega_{\rm gw}}(r,x) &= e^{+i (\omega_{\rm gw} t - m_{\rm gw} \varphi)}{}_{-2}\tau_{\mu\nu}t^{(g)\mu\nu}   ~.
\end{align}
Here, ${}_{s}\tau_{\mu\nu}$ is the adjoint of ${}_{s}\tau_{\mu\nu}^*$, which generates the source term of the Teukolsky equation. We show details of the derivation in App.~\ref{app:A}.
The amplitude $Z$ determines the radiated energy flux as
\begin{align}\label{dEGWdt}
	\frac{dE_{\rm GW}}{dt} &= \frac{1}{4\pi \omega_{\rm gw}^2}\sum_l \left|{}_{-2}Z_{lm_{\rm gw}\omega_{\rm gw}}\right|^2~.
\end{align}
From our normalization of the mode function, $dE_{\rm GW}/dt$ in \eqref{dEGWdt} corresponds to the energy flux computed classically with the cloud mass $\sim \mu$, {\it i.e.}, the occupation number $\sim 1$. Since radiative energy loss is weak compared with superradiant growth in the regime considered here~\cite{Yoshino:2013ofa}, we focus on the regime with $\omega_I t\gg 1$. 

In the following analysis, we only retain the dominant component in Eq.~\eqref{eq:fullsource}, namely the growing components.
We approximate the interaction Hamiltonian as
\begin{align}\label{eq:inthamilt}
	\hat H_{\rm int} &\sim - \frac{1}{4}\sum_{l,m,P}\int_{0}^{+\infty}d\omega \cr
    &\qquad \times \left\{C^{(g)}_{lm\omega} \hat{b}^{{\rm (in)}\dagger}_{lm\omega P} \hat{c}_g^2 e^{+i(\omega -\omega_{\rm gw})t + 2 \omega_I t} + {\rm h.c.}\right\}~.
\end{align}

\section{Decoherence from graviton emission}
\label{sec:4}

The interaction derived above correlates the cloud with the gravitons it emits. To describe the reduced dynamics of the cloud, we trace out the environmental gravitational degrees of freedom. We first examine interference between cloud configurations and then the evolution of the full two-mode covariance matrix.

\subsection{Master equation}
\label{sec:master}

We take the gravitons to be initially in the vacuum, uncorrelated with the cloud. We further assume the Born approximation, that is, the interaction is weak and the total density matrix can approximated by factorized form, $\rho_{\rm tot}(t) \sim \rho_I(t) \otimes \rho_\psi$. 
Here, $\rho_\psi$ is the density matrix of the graviton vacuum state.  
Following the standard open-system procedure~\cite{gardiner2004quantum} (see also Refs.~\cite{Oniga:2017pyq,Toros:2020krn} for nonrelativistic quantum matter coupled to gravitons), the evolution equation with leading order in $\hat H_{\rm int}$ is
\begin{align}
	\frac{d \rho_I}{d t} &\sim - \int_0^t ds\, {\rm Tr}_{\psi}\left\{[\hat H_{\rm int}(t), [\hat H_{\rm int}(s), \rho_I(s) \otimes \rho_{\psi}]]\right\}~.
\end{align}
We next replace $\rho_I(s)$ by $\rho_I(t)$, set $\tau=t-s$, and extend the upper limit of the $\tau$ integral to infinity, which is the so-called Markov approximation. We also neglect the Lamb shift for simplicity~\footnote{The Lamb shift introduce Hamiltonian of the form $\hat{H}_{\rm LS}= \lambda_{\rm LS} e^{4 \omega_I t}\hat{c}_g^{\dagger 2} \hat{c}_g^2$, where the constant $\lambda_{\rm LS}$ is determined by $|C^{(g)}_{lm\omega}|^2$. Including the Lamb shift would change the numerical factor in the analysis of Sec.~\ref{sec:covariance}. However, since we expect $\lambda_{\rm LS}$ to be same order as $\gamma_{\rm gw}$, our conclusion is not affected.}. The resulting master equation is 
\begin{align}\label{eq:masterint}
	\frac{d \rho_I}{d t} &= \frac{\gamma_{\rm gw}}{4}e^{4\omega_I t}\mathcal{D}[\hat{c}_g^2]\rho_I~,
\end{align}
where the superoperator $\mathcal{D}$ is defined as 
\begin{align}
	\mathcal{D}[\mathcal{O}]\rho &= \mathcal{O}\rho \mathcal{O}^\dagger - \frac{1}{2}\{\mathcal{O}^\dagger \mathcal{O},\rho \}~.
\end{align}
Here $\rho_I=U_0^\dagger\rho_SU_0$, with $U_0=e^{-i\hat H_0 t}$, and $\hat{c}_g$ is defined by the operators at the initial time (see Eq.~\eqref{eq:growdecay}). The coefficient $\gamma_{\rm gw}$ is related to the energy flux of the cloud as
\begin{align}\label{eq:gwrate}
	\gamma_{\rm gw} &= \frac{ \pi}{2} \sum_{l,P}|C^{(g)}_{l m_{\rm gw} \omega_{\rm gw} P} |^2 = \frac{1}{\omega_{\rm gw}} \frac{dE_{\rm GW}}{dt}~.
\end{align}
In the Schr\"odinger picture, the squeezing is carried by the state and the equation becomes
\begin{align}\label{eq:Schmaster}
	\frac{d \rho_S}{d t} &= -i [\hat{H}_0, \rho_S] + \frac{\gamma_{\rm gw}}{4}\mathcal{D}[\hat{c}_g^2]\rho_S~.
\end{align}

The Markov approximation assumes that the correlation time of the bath, which should be order of the light crossing time of the cloud, is short compared with the relevant evolution times of the cloud. The light crossing time of the cloud is around $t_{\rm cl}\sim \mu^{-1}/(G \mu M)$, which is much shorter than the growth time $\omega_I^{-1}$. Furthermore, before the cloud completely extract the angular momentum of the black hole, $\omega_I^{-1} \ll (\gamma_{\rm gw} N)^{-1}$ holds, where $N$ is the particle number of the cloud~\cite{Yoshino:2013ofa}. Therefore, the time scale of the backreaction due to the graviton emission and the growth by the superradiance is much slower than the bath correlation time and Markov approximation gives a good approximation for the superradiant cloud.

\subsection{Decoherence time}
\label{sec:decoherence}

We first estimate when graviton emission suppresses interference between the cloud states $\ket{c}$ and $\ket{c'}$, introduced in Eq.~\eqref{eq:cgeigenbasis}. 
Taking matrix elements of the master equation~\eqref{eq:masterint} in this basis gives
\begin{align}\label{eq:kernelrate}
    \frac{d}{dt} \braket{c'|\rho_I|c} &= -\frac{\gamma_{\rm gw}}{8}e^{4 \omega_I t}(|c|^4 + |c'|^4 - 2 {c'}^2 \bar{c}^2)\braket{c'|\rho_I|c}~,
\end{align}
which leads to
\begin{align}\label{eq:decoherencekernel}
    |\braket{c'|\rho_I|c}| \propto \exp\left(-\frac{\gamma_{\rm gw}}{32 \omega_I}(e^{4\omega_I t} - 1)|c^2 - {c'}^2|^2\right)~.
\end{align}
Equation~\eqref{eq:decoherencekernel} shows that interference between $\ket c$ and $\ket{c'}$ is suppressed when $c^2\neq {c'}^2$. 
Because the stress-energy tensor is quadratic in the scalar field $\phi$, the interference between $c$ and $-c$ remains unchanged.
The suppression is significant when the magnitude of the exponent is of order unity or larger. 

For the vacuum distribution~\eqref{eq:cgdistribution},
\begin{align}\label{eq:vaccgcg}
    \braket{\Omega(0)|\hat c_g^{\dagger 2}\hat c_g^2|\Omega(0)} &= 2~,
\end{align}
and the phase of $c$ is uniformly distributed. Thus, for two amplitudes drawn independently from this distribution, a typical separation is $|c^2-{c'}^2|^2\sim4$. Setting the magnitude of the exponent in Eq.~\eqref{eq:decoherencekernel} to order unity, we obtain the decoherence time in the regime $\omega_I t\gg1$ as
\begin{align}\label{eq:td}
    t_{\rm d} &= \frac{1}{4\omega_I} \log\left(\frac{8\omega_{\rm I}}{\gamma_{\rm gw}}\right)~.
\end{align}
The occupation number evaluated at $t=t_d$ in the free evolution is
\begin{align}\label{eq:numdecoh}
    N_d &= \sinh^2 \omega_I t_d \sim\sqrt{\frac{\omega_I}{2\gamma_{\rm gw}}}\gg 1~.
\end{align}
Using $N\sim e^{2\omega_I t}/4$, we find Eq.~\eqref{eq:decoherencekernel} indicates strong suppression of the off-diagonal elements of the density matrix for $|c^2-{c'}^2|\gg N_d/N$. 

The meaning of $t_d$ can be tied to the number of the emitted gravitons from the cloud. The expected number of gravitons emitted by time $t$ is
\begin{align}\label{eq:ngrav}
    n_{\rm grav}(t) = \int_0^t dt'\, \frac{\gamma_{\rm gw}}{4}\braket{\hat{c}_g^{\dagger 2}(t')\hat{c}_g^2(t')} = \frac{\gamma_{\rm gw}}{8 \omega_I}\left(e^{4 \omega_I t} - 1\right)~.
\end{align}
We used the fact that $\braket{\hat c_g^{\dagger 2}\hat c_g^2}$ is unchanged since $\hat{c}_g$ commutes with the interaction Hamiltonian. At $t=t_d$, we find $n_{\rm grav}(t_d)\sim1$. Thus, for the typical separation considered above, interference is significantly suppressed when the expected number of emitted gravitons exceeds order unity. A similar result has been found for decoherence of macroscopic superpositions of the point particle~\cite{Danielson:2024yru,Matsui:2026skx}.

Setting $c'=c$ in Eq.~\eqref{eq:kernelrate} gives $\partial_t\braket{c|\rho_I|c}=0$. Graviton emission therefore leaves the initial probability distribution $p_0(c)$ in Eq.~\eqref{eq:cgdistribution} unchanged. At a later time $t$, the growing part of the expectation values of the cloud operators are still obtained by drawing $c$ from the probability distribution $p_0(c)$ and evolving it according to Eq.~\eqref{eq:classicalgrowth}. The suppression of interference does not narrow the initial phase-space distribution.

\subsection{Evolution of the covariance matrix}
\label{sec:covariance}

We next calculate the moments $M_2$, $G_2$, and $Q$ defined in Eq.~\eqref{eq:M2G2Qdef} to determine how graviton emission changes the covariance matrix. For $M_2$, the master equation~\eqref{eq:Schmaster} gives
\begin{align}
	 \frac{dM_2}{dt}
    &=-2\omega_I M_2
    -\frac{\gamma_{\rm gw}}{2}M_4~,
\end{align}
where $M_4 = \braket{\hat{c}_g^{\dagger 2} \hat{c}_g \hat{c}_d + {\rm h.c.}}$. To evaluate the dissipative term, we use
\begin{align}\label{eq:disstrace}
	\mathrm{Tr}\left[\hat{O} \mathcal{D}[\hat{A}] \rho_S \right] &= \frac{1}{2}\mathrm{Tr}\left[(\hat{A}^\dagger[\hat{O},\hat{A}] + [\hat{A}^\dagger,\hat{O}]\hat{A}) \rho_S \right]~,
\end{align}
and the commutation relations~\eqref{eq:commutationc}. The same calculation gives
\begin{align}\label{eq:momenthierarchy}
    \frac{dM_4}{dt}
    &=2\omega_I M_4-\gamma_{\rm gw}M_6~,\\
    \frac{dM_6}{dt}
    &=6\omega_I M_6~,
\end{align}
with $M_6 = \braket{\hat{c}_g^{\dagger 3}\hat{c}_g^3}$. Under leaving the growing components in the interaction Hamiltonian~\eqref{eq:inthamilt}, the equation for $M_6$ contains no higher moments and the moment equation close at sixth order.

For the initial vacuum, we have
\begin{align}
    M_2(0) &=1~, &
    M_4(0) &=-4~, &
    M_6(0) &=6~,
\end{align}
and the solutions are
\begin{align}
    M_6(t)
    &=6e^{6\omega_I t}~,\\
    M_4(t)
    &=e^{2\omega_I t}
    \left[
    -4-\frac{3\gamma_{\rm gw}}{2\omega_I}
    \left(e^{4\omega_I t}-1\right)
    \right]~,\\
    M_2(t)
    &=e^{-2\omega_I t}
    \left[
    1+\frac{\gamma_{\rm gw}}{2\omega_I}
    \left(e^{4\omega_I t}-1\right)
    +\frac{3\gamma_{\rm gw}^2}{32\omega_I^2}
    \left(e^{4\omega_I t}-1\right)^2
    \right]~.
    \label{eq:M2solution}
\end{align}
We write the solution in terms of $N=\sinh^2(\omega_I t)$, the occupation number for free evolution. With graviton emission, the two occupations differ by $Q$, and $N$ approximates each occupation while corrections to growth are small. For $\omega_I t\gg1$, we have $N\sim e^{2\omega_I t}/4$, and combining with Eq.~\eqref{eq:numdecoh} gives
\begin{align}
    M_2
    \sim
    \frac{1}{4N}
    +\frac{N}{N_d^2}
    +\frac{3N^3}{2N_d^4}~.
    \label{eq:M2largeN}
\end{align}
The first term comes from the squeezing Hamiltonian $\hat H_S$ and decreases as the cloud grows. The remaining terms come from graviton emission and eventually make $M_2$ increase~\footnote{Although $\braket{\hat{c}_d^\dagger \hat{c}_d}$ grows, it is still much smaller than $\braket{\hat{c}_g^\dagger \hat{c}_g}$ for $N \ll N_d^2$, as one can check from Eqs.~\eqref{eq:M2largeN} and~\eqref{eq:G2grow}. Thus, leaving only the growing component in the interaction Hamiltonian~\eqref{eq:inthamilt} is justified for the analysis here.}.

Since $G_2 = \braket{\hat{c}_g^\dagger \hat{c}_g}$ and $[\hat c_g,\hat c_g^\dagger]=0$, 
the interaction with graviton (second term of the right-hand-side of Eq.~\eqref{eq:Schmaster}) does not change evolution of $G_2$.
Therefore, the evolution of $G_2$ is the same as for free evolution
\begin{align}\label{eq:G2grow}
	G_2(t)
	&=e^{2\omega_I t}~.
\end{align}
For the particle number difference $Q = \braket{\hat{n}_+ - \hat{n}_-}$, notice that 
\begin{align}
	[\hat{n}_+ - \hat{n}_-, \hat{c}_g^2] &= -2 \hat{c}_g^2~.
\end{align}
With Eq.~\eqref{eq:disstrace} and the fact particle number difference is preserved by the two-mode squeezing, we have
\begin{align}\label{eq:Qevol}
	\frac{dQ}{dt}
	&=
	-\frac{\gamma_{\rm gw}}{2}
	\braket{\hat{c}_g^{\dagger2}\hat{c}_g^2}
	=  -2 \dot{n}_{\rm grav}(t)~.
\end{align}
We use the emitted graviton number~\eqref{eq:ngrav}. The factor 2 comes from the fact that the graviton is created by the pair annihilation of the boson. 
Solving Eq.~\eqref{eq:Qevol} with the initial condition $Q(0)=0$ yields
\begin{align}
	Q(t)
	&= -2 n_{\rm grav}(t) =
	-\frac{\gamma_{\rm gw}}{4\omega_I}
	\left(e^{4\omega_I t}-1\right)~.
	\label{eq:Qsolution}
\end{align}

We substitute $M_2$, $G_2$, and $Q$ into Eq.~\eqref{eq:PPTnu} to evaluate the entanglement criterion. For $1\ll N\ll\omega_I/\gamma_{\rm gw}$, both $M_2/G_2$ and $|Q|/G_2$ are small. Expanding in these ratios gives
\begin{align}
	\widetilde\nu_-
	&\sim
	\frac{1}{8N}
	+\frac{N}{2N_d^2}
	+\frac{N^3}{4N_d^4}~.
	\label{eq:PPTlargeN}
\end{align}
The squeezing first decreases $\tilde{\nu}_-$, while graviton emission eventually makes it increase. When $\tilde{\nu}_-$ reaches $1/2$, the covariance matrix no longer certifies entanglement (see Eq.~\eqref{eq:PPTbound}). The corresponding value of $N$ is
\begin{align}
	N_{\rm ent}
	&\sim
	2^{1/3}N_d^{4/3}~.
	\label{eq:Ncov}
\end{align}

After $N \gtrsim N_{\rm ent}$, the first and second moments can be reproduced by a statistical mixture of coherent states~\cite{Duan:2000awi}. For the covariance matrix in Eq.~\eqref{eq:covblock}, $\widetilde\nu_-$ is also the smallest ordinary eigenvalue of $V$.~\footnote{Since partial transposition reverses the sign of $\hat p_-$ and preserves the ordinary eigenvalues of the covariance matrix, $\widetilde\nu_-$ is also the smallest eigenvalue of $V$.} Once $\tilde\nu_-\geq1/2$, the matrix $V-I/2$ is positive semidefinite. 
In this regime, all first and second moments can be reproduced by density matrix with the statistical mixture of coherent states
\begin{align}
	\rho_{\rm cl}(t) = \int \frac{d^2 \alpha_+ d^2 \alpha_-}{\pi^2} P(\alpha_+,\alpha_-;t) \ket{\alpha_+,\alpha_-}\bra{\alpha_+,\alpha_-}~,
\end{align}
where the probability distribution is given by
\begin{align}
	P(\alpha_+,\alpha_-;t) &= \frac{1}{\sqrt{\det \Sigma}}\exp\left(-\bm{\alpha}\cdot \Sigma^{-1}(t) \cdot \bm{\alpha}\right)~, 	
\end{align}
with $\Sigma(t) = V(t) - \frac{1}{2}I$ and $\bm{\alpha} = (\alpha_{+,R},\alpha_{-,R},\alpha_{+,I},\alpha_{-,I})$.
Therefore, measurements of the first and second moments alone cannot deny the cloud being such a statistical mixture. However, the cloud becomes non-Gaussian under the non-linear dissipation $\mathcal{D}[\hat c_g^2]$, so determining whether the non-classicality remains after $N \gtrsim N_{\rm ent}$ requires higher moments or the full density matrix~\cite{Shchukin:2005slb,Miranowicz:2009qry}, which we leave for future work.

\subsection{Comparison of the scales}
\label{sec:scales}

We compare $N_d$ and $N_{\rm ent}$ with the occupation numbers at which the classical growth of the cloud is significantly modified, 
which can be characterized by two relevant scales. First, superradiant growth eventually stops because the cloud extract momentum from the black hole and spins it down until it do not have enough angular momentum to support superradiance. 
If other mechanisms that stop the growth are neglected, angular momentum conservation gives an estimate of the occupation number at spin-down saturation~\cite{Arvanitaki:2014wva}
\begin{align}\label{eq:Nstar}
    N_* &= \frac{M^2}{m_c}\Delta \chi~,
\end{align}
where $\Delta\chi$ is the change in the dimensionless black hole spin required for the superradiant condition to cease to hold.

The second scale is reached when the gravitational wave energy loss becomes comparable to the energy supplied by superradiant growth. We define this occupation number by 
\begin{align}\label{eq:Nbal}
    N_{\rm bal} &\sim\frac{\omega_I}{\gamma_{\rm gw}}~,
\end{align}
up to a numerical factor. For $N\ll N_{\rm bal}$, radiative energy loss has little effect on the cloud growth.

For $\mu M\ll1$, the graviton emission rate and the superradiant growth rate are approximated by
\begin{align}
	\gamma_{\rm gw} &\sim C_{nl} ( \mu M)^{4 l + 11} \left(\frac{M_{\rm pl}}{M}\right)^3 M_{\rm pl}~,\\
	\omega_I &\sim D_{nl}(m\Omega_H-\omega_R)(\mu M)^{4l+5}~,
\end{align}
where $C_{nl}$ and $D_{nl}$ depend on the cloud quantum numbers $(n,l,m)$ (see, for example, Refs.~\cite{Brito:2015oca,Yoshino:2013ofa}).  As an example, the occupation scales for the fastest growing mode $n =2, l=m=1$ are estimated as
\begin{align}
	N_d & \sim 10^{41} \left(\frac{ \mu M}{0.1} \right)^{-3} \left(\frac{M}{M_{\odot}}\right)~,\cr
	N_{\rm ent} & \sim 10^{55} \left(\frac{ \mu M}{0.1} \right)^{-4} \left(\frac{M}{M_{\odot}}\right)^{4/3}~,\cr
	N_* &  \sim 10^{75}\left(\frac{\Delta \chi}{0.1} \right) \left(\frac{M}{M_{\odot}}\right)^2~,\cr
	N_{\rm bal} &\sim 10^{82}\left(\frac{ \mu M}{0.1} \right)^{-6} \left(\frac{M}{M_{\odot}}\right)^2~.\label{eq:Ndnum}
\end{align}
 For the parameters illustrated in Eq.~\eqref{eq:Ndnum}, the scales then satisfy
\begin{align}\label{eq:hierarchy}
	1 \ll N_d \ll N_{\rm ent} \ll N_* \ll N_{\rm bal}~.
\end{align}
Thus, the suppression of the interference between the different growing mode amplitude basis ($N\sim N_d$) and the loss of the entanglement certification by the covariance matrix ($N\sim N_{\rm ent}$) occur much before the spin-down saturation ($N \sim N_*$). Furthermore, they occur when the effect of the energy loss due to graviton on the growth of the cloud is small.

\section{Conclusion}
\label{sec:5}

A superradiant cloud growing from vacuum fluctuations remains in a two-mode squeezed state under free evolution, with strong entanglement between two modes even at large occupation number. We have studied how graviton emission changes the state of the cloud by deriving a master equation by tracing out gravitons radiated to infinity. We solve the master equation under the vacuum initial condition, keeping the Kerr background fixed and retaining only the growing mode of the cloud in the interaction Hamiltonian.

We showed that graviton emission suppresses quantum interference between configurations with different cloud amplitudes. For typical amplitudes drawn from the vacuum distribution, the suppression becomes significant when the expected number of emitted gravitons exceeds order unity. The dissipative term we considered here does not affect the amplitude distribution. The amplitude and phase distributions of the cloud inherit the initial broad distribution determined by the quantum fluctuation, and the amplitude is simply magnified by the superradiant instability.

We also compute the evolution of the covariance matrix to see whether the entanglement generated by the superradiance remains in the covariance matrix. The covariance matrix still certifies entanglement when quantum interference starts to be suppressed. However, at a sufficiently large occupation number, all first and second moments become able to be reproduced by a classical statistical mixture of the two-mode coherent states. Then, measurements of these moments alone can no longer distinguish the cloud from such a statistical mixture. For the astrophysical parameters, both changes occur well before spin-down saturation, while energy loss due to graviton emission has little effect on cloud growth. Thus, graviton emission suppress quantum interference between the state with different amplitudes of the growing modes and 
erase the entanglement signature accessible by the first and the second moment while the occupation numbers continue to grow approximately as in the free evolution.

Our computation alone cannot tell about the state near the spin-down saturation. Describing the cloud near saturation requires allowing the black hole spin, and hence the growth rate, to change. The importance of the ignored decaying mode in the interaction should also be considered in this regime. Other than the spin-down, the self-interactions may excite other cloud modes and introduce additional dissipation~\cite{Mocanu:2012fd, Baryakhtar:2020gao, Omiya:2022gwu}. How these effects modify the decoherence scales and the subsequent cloud state remains to be determined.

The contribution of gravitons associated with horizon flux also remains to be evaluated. Horizon-induced decoherence has been studied for superposition of the point particles near a black hole~\cite{Danielson:2022tdw,Gralla:2023oya}. We leave extending our analysis to include such effects to future work.

\section*{Acknowledgement}
We thank Xi Tong and Yi Wang for discussion during the early part of the project.
H. O. was supported by JSPS KAKENHI Grant Numbers JP23H00110 and JP25K17388. L. F. would like to thank Jun Zhang from ICTP-AP, Beijing, for his support. H.Y.Z. is supported by IBS under project code IBS-R018-D3. T.~T.~is supported by Grant-in-Aid for Scientific Research under Contract No.~JP23H00110, and also by SPIRIT2 2026 of Kyoto University. 

\appendix

\section{Derivation of the flux}\label{app:A}

In this appendix, we work at leading order in $\omega_I/\omega_R$, {\it i.e.}, neglecting the slowly varying growth factor $e^{2\omega_I t}$ when performing the time integral.
Rather than evaluating Eq.~\eqref{eq:coefCg} at fixed time, we first integrate it over time,
\begin{align}
	A &= \int d^4 x \sqrt{-g}\, \bar{\overleftarrow{\Pi}}_{\mu\nu}(x;l m \omega P) t^{(g)\mu\nu}\cr
	&= 2\pi C^{(g)}_{lm\omega} \delta_{m m_{\rm gw}}\delta(\omega - \omega_{\rm gw})~.
\end{align}
We decompose $A$ as
\begin{align}
    A &=   I + P J~,
\end{align}
with
\begin{align}
	I &= \int dt' \int d r' dx' d\varphi' {\rho'}^{2}  \overline{{}_{-2}\overleftarrow{R}_{lm\omega}}(r') \overline{{}_{+2}S_{lm\omega}}(x')\cr
	&\qquad  \times  \frac{e^{+i(\omega t' - m \varphi')}}{\sqrt{2\pi}} {}_{+2}\tau_{\mu\nu}t^{(g)\mu\nu}(x')~,\\
	J &= \int dt' \int d r' dx' d\varphi' {\rho'}^{2}     \overline{{}_{-2}{\overleftarrow{R}}_{lm\omega}}(r') \overline{{}_{-2}S_{lm\omega}}(x')\cr
	&\qquad  \times  \frac{e^{+i(\omega t' - m \varphi')}}{\sqrt{2\pi}} \overline{{}_{+2}\tau_{\mu\nu}}t^{(g)\mu\nu}(x')~,
\end{align}
where ${\rho'}^{2}  = {r'}^2 + a^2 {x'}^2$.
Here, we use the definition of the adjoint operator
\begin{align}
	\int d^4 x \sqrt{-g}T_{\mu\nu}\, \overline{{}_{+2}\tau^{*\mu\nu}X} &= \int d^4 x \sqrt{-g}\bar{X}\, {}_{+2}\tau^{\mu\nu}T_{\mu\nu}~,
\end{align}
for an arbitrary scalar function $X$.

To evaluate $I$, first flip the signs of $t'$ and $\varphi'$ simultaneously. The operator then transforms as ${}_{+2}\tau_{\mu\nu}\to4\Delta^{-2}{}_{-2}\tau_{\mu\nu}$.
With an additional flip $(\omega, m) \to (-\omega, -m)$ and the relation~\cite{Sago:2005fn}
\begin{align}
	\overline{{}_{-2}R_{l(-m)(-\omega)}}(r) &= {}_{-2}R_{lm\omega}(r)~,\cr
	\overline{{}_{+2}S_{l(-m)(-\omega)}}(x) &= (-1)^m {}_{-2}S_{lm\omega}(x)~,
\end{align}
we have
\begin{align}
	I
	&= 4(-1)^m\int dr dx\,(r^2+a^2x^2)\Delta^{-2}
	{}_{-2}\overleftarrow{R}_{lm\omega}(r){}_{-2}S_{lm\omega}(x)\cr
	&\qquad\times\int_{-\infty}^{+\infty}dt\int_0^{2\pi}d\varphi\,
	\frac{e^{+i(\omega t-m\varphi)}}{\sqrt{2\pi}}
	{}_{-2}\tau_{\mu\nu}t^{(g)\mu\nu}\cr
	&= \frac{(-1)^m}{2\pi i \omega \sqrt{4\pi \omega}} {}_{-2}Z_{lm_{\rm gw}\omega_{\rm gw}} (2\pi)^{3/2}\delta_{mm_{\rm gw}}\delta(\omega - \omega_{\rm gw})~.
\end{align}
A similar calculation, with the additional transformation $x\to-x$, gives
\begin{align}
	J &= (-1)^{l} I ~.
\end{align}
Here, we use the transformation property
\begin{align}
	{}_{-2}S_{lm\omega}(-x) &= (-1)^l {}_{+2}S_{lm\omega}(x)~.
\end{align}
To summarize, we have the relation
\begin{align}
	C_{lm\omega P } &= \frac{(-1)^m}{i\pi (2 \omega)^{3/2}}{}_{-2}Z_{lm_{\rm gw}\omega_{\rm gw}}\left(1 + P (-1)^l\right)
\end{align}
For even ($P = +1$) and odd parity ($P = -1$) mode, only modes with $l=2,4,\cdots$ and $l= 3, 5, \cdots$ are emitted, respectively~\cite{Yoshino:2013ofa}.

\bibliography{axionref}

\end{document}